\documentclass[twoside,a4paper,11pt]{proceedings}
\usepackage{graphicx}
\usepackage{hyperref}
\usepackage{movie15}
\usepackage{natbib}
\def\hoy{\number\day \space de \space\ifcase\month\or
 Enero\or Febrero\or Marzo\or Abril\or Mayo\or Junio\or
 Julio\or Agosto\or Septiembre\or Octubre\or Noviembre\or Diciembre\fi
 \space de \number\year}
\def\ii/{\'{\i}}
\def\cion/{ci\'on}
\def\cao/{\c c\~ao}
\def\arcsec{\hbox{$^{\prime\prime}$}}
\def\utw{\smash{\rlap{\lower5pt\hbox{$\sim$}}}}
\def\udtw{\smash{\rlap{\lower6pt\hbox{$\approx$}}}}

\def\farcs{\hbox{$.\!\!^{\prime\prime}$}}

\def\tens#1{\ifmmode\mathchoice{\mbox{$\sf\displaystyle#1$}}
{\mbox{$\sf\textstyle#1$}}
{\mbox{$\sf\scriptstyle#1$}}
{\mbox{$\sf\scriptscriptstyle#1$}}\else
\hbox{$\sf\textstyle#1$}\fi}
\def\vec#1{\ifmmode\mathchoice{\mbox{\boldmath$\displaystyle#1$}}
{\mbox{\boldmath$\textstyle#1$}}
{\mbox{\boldmath$\scriptstyle#1$}}
{\mbox{\boldmath$\scriptscriptstyle#1$}}\else
\hbox{\boldmath$\textstyle#1$}\fi}
\def\bbbc{{\mathchoice {\setbox0=\hbox{$\displaystyle\rm C$}\hbox{\hbox
to0pt{\kern0.4\wd0\vrule height0.9\ht0\hss}\box0}}
{\setbox0=\hbox{$\textstyle\rm C$}\hbox{\hbox
to0pt{\kern0.4\wd0\vrule height0.9\ht0\hss}\box0}}
{\setbox0=\hbox{$\scriptstyle\rm C$}\hbox{\hbox
to0pt{\kern0.4\wd0\vrule height0.9\ht0\hss}\box0}}
{\setbox0=\hbox{$\scriptscriptstyle\rm C$}\hbox{\hbox
to0pt{\kern0.4\wd0\vrule height0.9\ht0\hss}\box0}}}}
\def\bbbq{{\mathchoice {\setbox0=\hbox{$\displaystyle\rm
Q$}\hbox{\raise
0.15\ht0\hbox to0pt{\kern0.4\wd0\vrule height0.8\ht0\hss}\box0}}
{\setbox0=\hbox{$\textstyle\rm Q$}\hbox{\raise
0.15\ht0\hbox to0pt{\kern0.4\wd0\vrule height0.8\ht0\hss}\box0}}
{\setbox0=\hbox{$\scriptstyle\rm Q$}\hbox{\raise
0.15\ht0\hbox to0pt{\kern0.4\wd0\vrule height0.7\ht0\hss}\box0}}
{\setbox0=\hbox{$\scriptscriptstyle\rm Q$}\hbox{\raise
0.15\ht0\hbox to0pt{\kern0.4\wd0\vrule height0.7\ht0\hss}\box0}}}}
\def\bbbt{{\mathchoice {\setbox0=\hbox{$\displaystyle\rm
T$}\hbox{\hbox to0pt{\kern0.3\wd0\vrule height0.9\ht0\hss}\box0}}
{\setbox0=\hbox{$\textstyle\rm T$}\hbox{\hbox
to0pt{\kern0.3\wd0\vrule height0.9\ht0\hss}\box0}}
{\setbox0=\hbox{$\scriptstyle\rm T$}\hbox{\hbox
to0pt{\kern0.3\wd0\vrule height0.9\ht0\hss}\box0}}
{\setbox0=\hbox{$\scriptscriptstyle\rm T$}\hbox{\hbox
to0pt{\kern0.3\wd0\vrule height0.9\ht0\hss}\box0}}}}
\def\bbbs{{\mathchoice
{\setbox0=\hbox{$\displaystyle     \rm S$}\hbox{\raise0.5\ht0\hbox
to0pt{\kern0.35\wd0\vrule height0.45\ht0\hss}\hbox
to0pt{\kern0.55\wd0\vrule height0.5\ht0\hss}\box0}}
{\setbox0=\hbox{$\textstyle        \rm S$}\hbox{\raise0.5\ht0\hbox
to0pt{\kern0.35\wd0\vrule height0.45\ht0\hss}\hbox
to0pt{\kern0.55\wd0\vrule height0.5\ht0\hss}\box0}}
{\setbox0=\hbox{$\scriptstyle      \rm S$}\hbox{\raise0.5\ht0\hbox
to0pt{\kern0.35\wd0\vrule height0.45\ht0\hss}\raise0.05\ht0\hbox
to0pt{\kern0.5\wd0\vrule height0.45\ht0\hss}\box0}}
{\setbox0=\hbox{$\scriptscriptstyle\rm S$}\hbox{\raise0.5\ht0\hbox
to0pt{\kern0.4\wd0\vrule height0.45\ht0\hss}\raise0.05\ht0\hbox
to0pt{\kern0.55\wd0\vrule height0.45\ht0\hss}\box0}}}}
\def\bbbz{{\mathchoice {\hbox{$\sf\textstyle Z\kern-0.4em Z$}}
{\hbox{$\sf\textstyle Z\kern-0.4em Z$}}
{\hbox{$\sf\scriptstyle Z\kern-0.3em Z$}}
{\hbox{$\sf\scriptscriptstyle Z\kern-0.2em Z$}}}}
\def\diameter{{\ifmmode\mathchoice
{\ooalign{\hfil\hbox{$\displaystyle/$}\hfil\crcr
{\hbox{$\displaystyle\mathchar"20D$}}}}
{\ooalign{\hfil\hbox{$\textstyle/$}\hfil\crcr
{\hbox{$\textstyle\mathchar"20D$}}}}
{\ooalign{\hfil\hbox{$\scriptstyle/$}\hfil\crcr
{\hbox{$\scriptstyle\mathchar"20D$}}}}
{\ooalign{\hfil\hbox{$\scriptscriptstyle/$}\hfil\crcr
{\hbox{$\scriptscriptstyle\mathchar"20D$}}}}
\else{\ooalign{\hfil/\hfil\crcr\mathhexbox20D}}%
\fi}}
\def\sq{\ifmmode\squareforqed\else{\unskip\nobreak\hfil
\penalty50\hskip1em\null\nobreak\hfil\squareforqed
\parfillskip=0pt\finalhyphendemerits=0\endgraf}\fi}
\def\squareforqed{\hbox{\rlap{$\sqcap$}$\sqcup$}}
\begin{document}
\pagenumbering{arabic}
\pagestyle{myheadings}
\thispagestyle{empty}
\vspace*{-1cm}
{\flushleft\includegraphics[width=8cm,viewport=0 -30 200 -20]{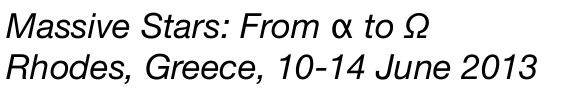}}
\vspace*{0.2cm}
\begin{flushleft}
{\bf {\LARGE
The Galactic O-Star Spectroscopic \linebreak
Catalog (GOSC) and Survey (GOSSS): \linebreak
first whole-sky results and further updates
}\\
\vspace*{1cm}
J. Ma\'{\i}z Apell\'aniz$^1$,
A. Sota$^1$,
N. I. Morrell$^2$,
R. H. Barb{\'a}$^3$,
N. R. Walborn$^4$, \linebreak
E. J. Alfaro$^1$,
R. C. Gamen$^5$,
J. I. Arias$^3$,
and
A. T. Gallego Calvente$^1$
%
}\\
\vspace*{0.5cm}
%
$^{1}$
Instituto de Astrof\'{\i}sica de Andaluc\'{\i}a-CSIC, Granada, Spain \\
$^{2}$
Las Campanas Observatory, La Serena, Chile \\
$^{3}$
Departamento de F\'{\i}sica, Universidad de La Serena, La Serena, Chile \\
$^{4}$
Space Telescope Science Institute, Baltimore, MD, USA \\
$^{5}$
Instituto de Astrof\'{\i}sica de La Plata (CONICET, UNLP), La Plata, Argentina
%
\end{flushleft}
\markboth{
GOSC and GOSSS: first whole-sky results and further updates
}{
Ma\'{\i}z Apell\'aniz et al.
}
\thispagestyle{empty}
\vspace*{0.4cm}
\begin{minipage}[l]{0.09\textwidth}
\ 
\end{minipage}
\begin{minipage}[r]{0.9\textwidth}
\vspace{1cm}
\section*{Abstract}{\small
The Galactic O-Star Spectroscopic Survey (GOSSS) is obtaining high quality $R\sim 2500$ blue-violet 
spectroscopy of all Galactic stars ever classified as of O type with $B < 12$ and a significant fraction of 
those with $B = 12-14$. As of June 2013, we have obtained, processed, and classified 2653 spectra of 1593 
stars, including all of the sample with $B < 8$ and most of the sample with $B = 8-10$, making GOSSS already 
the largest collection of high quality O-star optical spectra ever assembled by a factor of 3. We  
discuss the fraction of false positives (stars classified as O in
previous works that do not belong to that class) and the implications of the observed 
magnitude distribution for the spatial distribution of massive stars and dust within a 
few kpc of the Sun. We also present new spectrograms for some of the interesting objects 
in the sample and show applications of GOSSS data to the study of the intervening ISM. Finally, we 
present the new version of the Galactic O-Star Catalog (GOSC), which incorporates the data in GOSSS-DR1, and
we discuss our plans for MGB, an interactive spectral classification tool for OB stars.
\vspace{10mm}
\normalsize}
\end{minipage}

\begin{figure}
\center
\includegraphics[width=\textwidth,viewport=58 38 536 331]{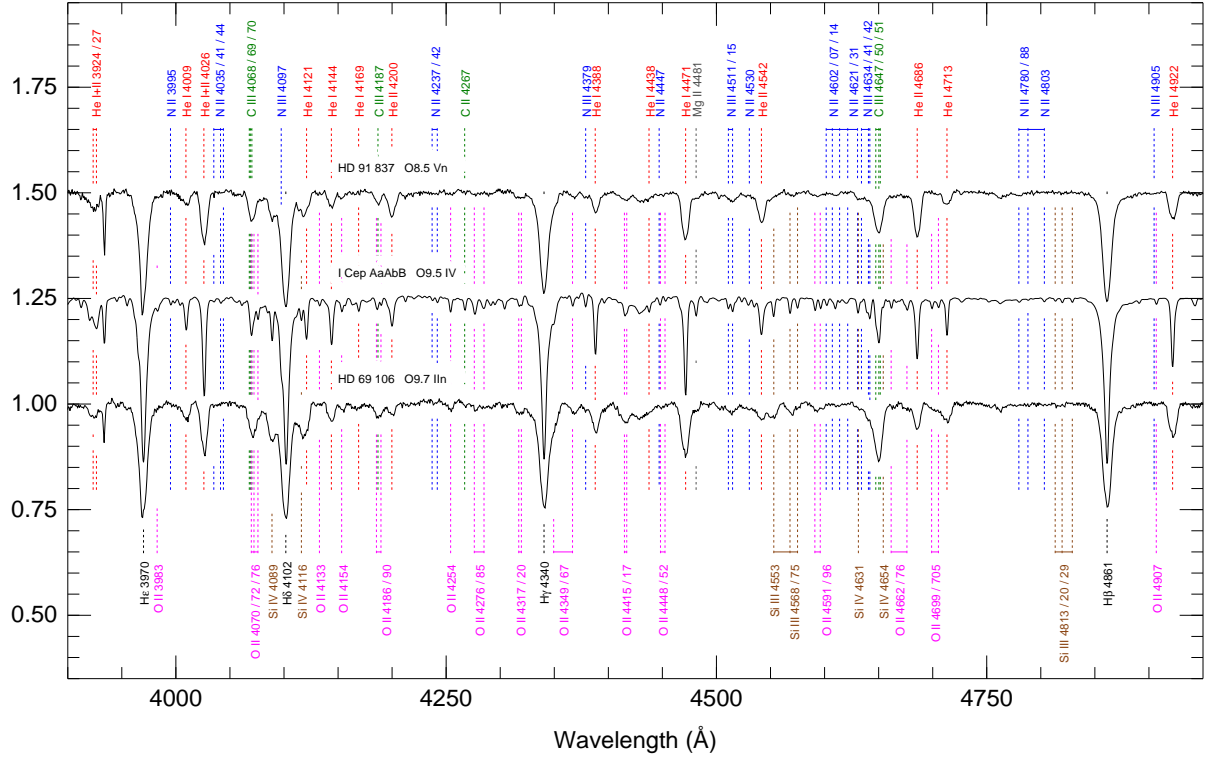} 
\caption{Three ``new'' bright O stars found by GOSSS.}
\end{figure}

\section{A history of the project}

\begin{itemize}
 \item 2004: v1.0 of the Galactic O-Star Catalog (GOSC, \citealt{Maizetal04b}).
  \begin{itemize}
   \item Literature spectral types (mostly 1970-80s by N. R. Walborn).
   \item Additional data from different sources. 
  \end{itemize}
 \item 2006: v2.0 of GOSC with many more stars \citep{Sotaetal08}. 
 \item 2007: GOSSS starts based on GOSC. 
  \begin{itemize}
   \item $R \sim 2500$, uniform quality blue-violet spectroscopy.
   \item Both hemispheres, complete to a given B magnitude.
   \item Revised spectral types for Galactic O stars.
  \end{itemize}
 \item 2010: First letter on Ofc and Of?p stars \citep{Walbetal10a}. 
 \item 2011: First paper with bright northern stars, complete set of standards, and revised classification criteria 
             (\citealt{Sotaetal11a}, paper I).
 \item 2011: Survey description \citep{Maizetal11}.
 \item 2012: Second letter on ONn stars \citep{Walbetal11}.
 \item 2012: GOSSS data used for NGC 1624-2 \citep{Wadeetal12b} and HD 120\,678 \citep{Gameetal12} papers.
\end{itemize}

\begin{figure}
\center
\includegraphics[width=\textwidth,viewport=58 38 536 331]{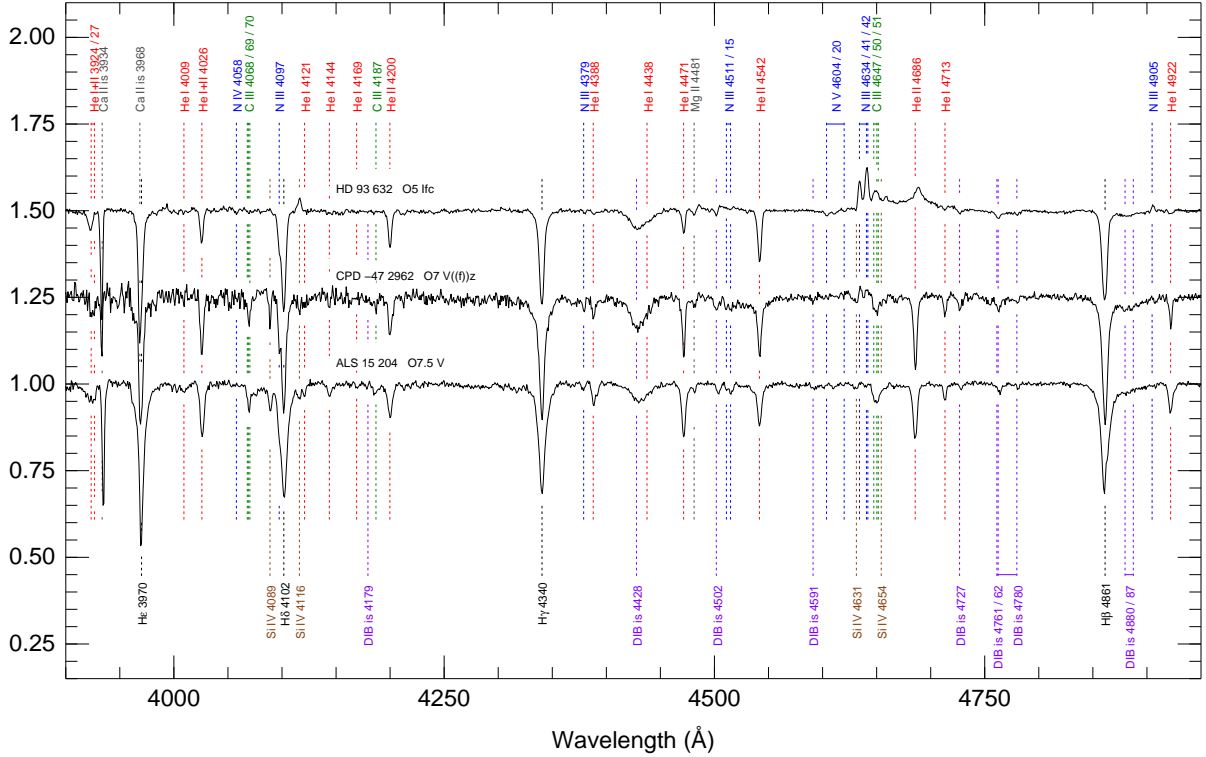} 
\caption{Three O stars without nearby companions in GOSSS-DR1.}
\end{figure}

\section{What is new?}
	
\begin{itemize}
 \item A second paper with bright southern stars is about to be submitted (Sota et al. 2013b).
 \item Spectral types and spectra for all the stars in the first two papers have been made public as 
       Data Release 1 (GOSSS-DR1, Sota et al. 2013a in these proceedings).
 \item GOSC has been remodeled (v3.0) to include the above.
 \item First whole-sky results have been obtained.
 \item GOSSS is now being used to study the Diffuse Interstellar Bands (DIBs) in the intervening ISM 
       \citep{Maizetal13a}.
 \item MGB, the accompanying spectral classification software \citep{Maizetal12}, will be released with 
       GOSSS-DR2.
\end{itemize}

\begin{figure}
\center
\includegraphics[width=\textwidth,viewport=58 38 536 331]{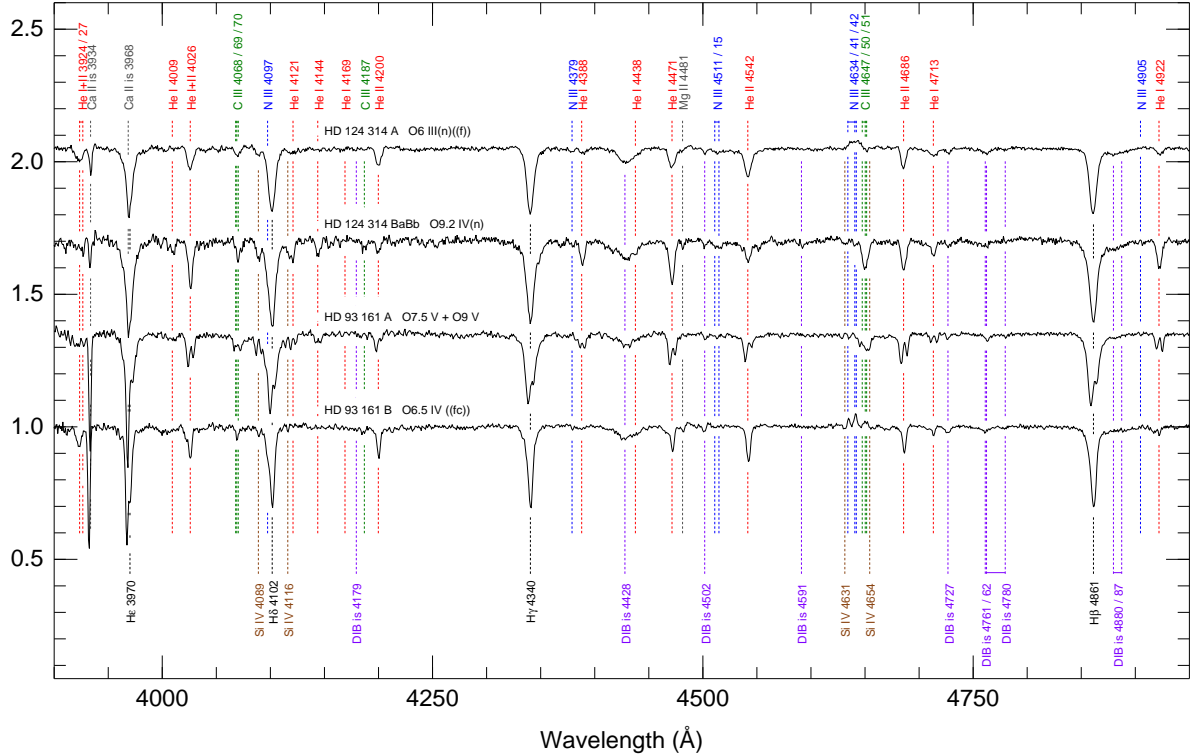} 
\caption{Examples of visual multiples in GOSSS-DR1.}
\end{figure}

\vfill

\eject

\section{GOSC v3.0}

\begin{itemize}
 \item Now based on GOSSS-DR1:
  \begin{itemize}
   \item 449 stars with spectral types from GOSSS papers I and II.
   \item Alternate and previous spectral types as additional columns. 
   \item Revised approximate $B$ and $J$ photometry ($B_{\rm ap}$, $J_{\rm ap}$) for all stars. Most of the
         $J_{\rm ap}$ values are taken from 2MASS with a few exceptions. The $B_{\rm ap}$ values are either
         Tycho-2 or Johnson and are, in general, less uniform than the $J_{\rm ap}$ ones. All the values have 
         been rounded up to one tenth of a magnitude in order not to overestimate their accuracy. In those cases
         where the WDS catalog \citep{Masoetal01} lists more than one visual component, the $B_{\rm ap}$ and 
         $J_{\rm ap}$ photometry refers to the brightest component included in the spectrum (with the flux 
         partition derived from the WDS $\Delta m$). This photometry can be used to estimate the color excess
         $E(B-J)$ knowing that $(B-J)_0$ ranges between $-0.7$ and $-1.0$ for O stars. $E(B-J)$ has the advantage
         over $E(B-V)$ of exploiting a longer baseline in wavelength (thus being more accurate when using diverse
         sources) and is preferred over $E(B-K)$ because $K$-band photometry may be affected by IR excesses. Also, 
         $E(B-J) \sim A_V$ for the most common extinction laws.
   \item Revised WDS \citep{Masoetal01} membership.
   \item Direct access to GOSSS spectra (FITS tables).
  \end{itemize}
 \item Future:
  \begin{itemize}
   \item New GOSSS data releases: O, WR, and other early-type stars: $\sim$2500 objects planned by 2016.
   \item LMC+SMC extension.
   \item CHORIZOS-derived SEDs \citep{Maiz04c} using the new family of extinction laws of \citet{Maiz13b}.
   \item Additional interface improvements.
  \end{itemize}
 \item Visit \url{http://gosc.iaa.es}.
\end{itemize}

\begin{figure}
\center
\includegraphics[width=\textwidth,viewport=58 38 536 331]{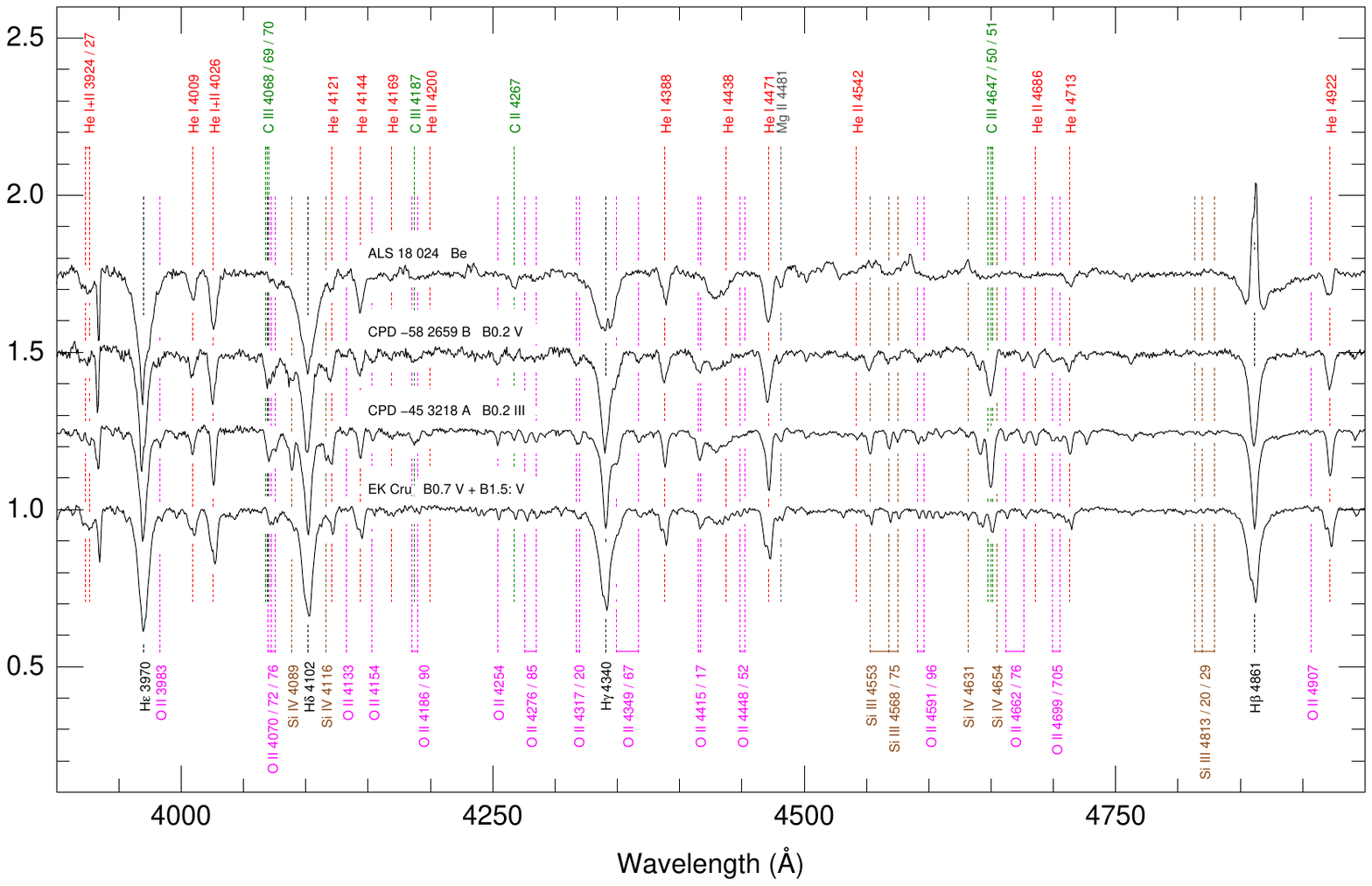} 
\caption{Four examples of false positives: B stars that were previously classified as being of O type by
\citet{MuzzMcCa73}, \citet{Feinetal80}, \citet{Garretal77}, and \citet{Garmetal82}, respectively.}
\end{figure}

\begin{figure}
\center
\includegraphics[width=\textwidth,viewport=58 38 536 331]{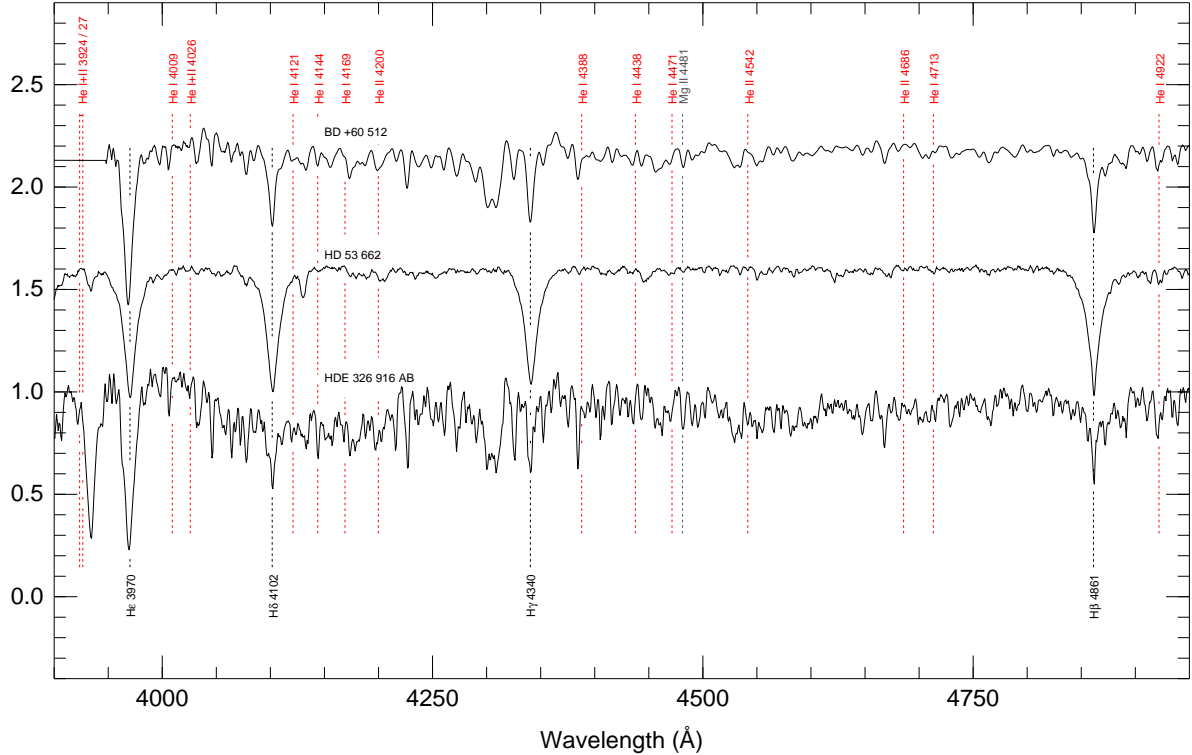} 
\caption{Three examples of false positives: stars that were previously classified as being of O type by 
\citet{Hilt56}, \citet{ContAlsc71}, and \citet{Schietal71}, respectively. These are likely confusions with 
BD~+60~513, HD~54\,662, and HDE~325\,516, respectively.}
\end{figure}

\vfill

\eject

\section{First whole-sky results of GOSSS}

\subsection{New spectral classifications}

\begin{itemize}
 \item They include all bright Galactic O stars (those with $B_{\rm ap} \le 8.0$) and dimmer objects. 
 \item Some bright O stars are ``new'' in the sense that most or all of the previous classifications were of B
       type (see Figure~1). For example, I~Cep~AaAbB (HD~202\,214~AaAbB), O9.5 IV, is likely a composite spectrum, 
       since it has three visual bright components (unresolved in GOSSS-DR1). It has at least ten previous 
       spectral classifications as B0 (including \citealt{Morgetal53b} and \citealt{Lesh68}) but we have 
       found only one as O9 \citep{MannHumb55}.
 \item HD~93\,632 was classified as O5~III~(f)var by \citet{Walb73a} but in Figure~2 it clearly has a 
       luminosity class of I, based on He~{\sc ii}~$\lambda$4686 strongly in emission. Recent OWN spectra 
       \citep{Barbetal10} show a weaker He~{\sc ii}~$\lambda$4686, indicating unusual variability, possibly 
       related to the presence of a magnetic field. 
 \item CPD~$-$47~2962 is one of many new members of the O Vz class discovered with GOSSS (Figure~2). 
 \item ALS~15\,204 is a serendipitous discovery: it was placed on the slit because another O star was observed
       nearby (Figure~2). A weak redward component may indicate the presence of a dim spectroscopic B 
       companion.
 \item HD~124\,314~A and BaBb are separated by 2\farcs5 and both have O spectral types (Figure~3). Ba and Bb 
       cannot be separated in GOSSS-DR1 as they are only 0\farcs21 apart. A is a likely SB system according to 
       preliminary OWN results \citep{Barbetal10} but is not seen as such here.
 \item HD~93\,161~A and B are separated by 2\arcsec\ and both are O-type systems (Figure~3). HD~93\,161~A 
       itself is an O+O spectroscopic binary.
\end{itemize}

\begin{figure}
\center
\includegraphics[width=0.7\textwidth,viewport=58 38 536 331]{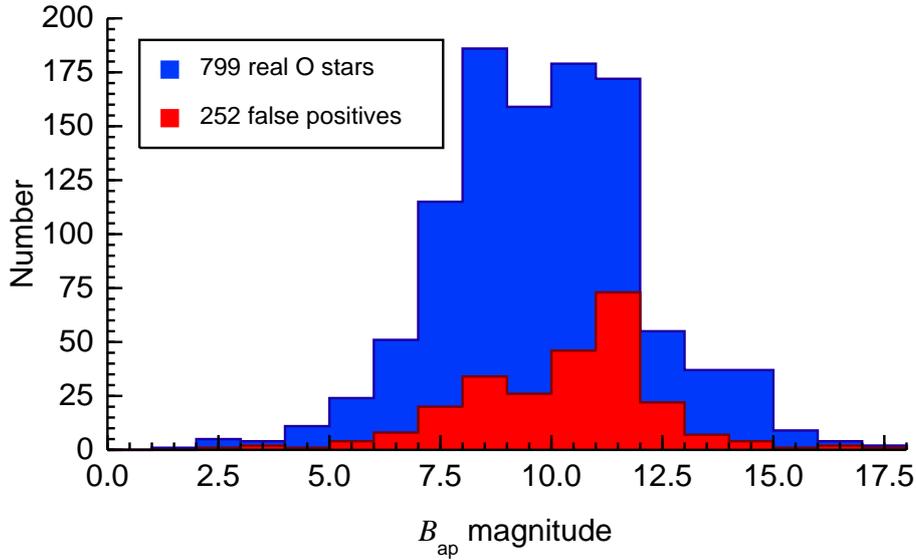} 
\caption{Real O stars and false positives currently observed with GOSSS as a function of $B_{\rm ap}$. 
For $B_{\rm ap} < 10.0$ the ratio of the first to the second is 460/96 = 4.79. For $B_{\rm ap} \ge 10.0$, the
ratio is 339/156 = 2.17, a sign that literature classifications become more unreliable for dimmer stars, as
expected.}
\end{figure}

\subsection{Lies, damn lies, and statistics}

$\,\!$\indent We have observed 1593 stars, which in the literature appeared as:

\begin{itemize}
 \item O: 1014.
 \item B or OB: 382.
 \item Other: 93.
 \item No previous classification: 104.
\end{itemize}

Of those 1593 stars, 799 turned out to be real O stars. The false positives (objects that went from O to non-O) 
were 252, yielding a false positive rate of 252/1014 = 24.9\%. Examples are found in Figures~4~and~5. 
Figure~6 shows the dependence with $B$ magnitude. The false positives were of spectral type:

\begin{itemize}
 \item A: 4.
 \item B: 213.
 \item F: 14.
 \item G: 5.
 \item K: 7.
 \item LBV: 1.
 \item PN: 4.
 \item sd: 4.
\end{itemize}

The false negatives (objects that went from non-O to O) were 37, yielding a false negative rate of 37/579 =
6.4\%. Those stars were previously classified as:

\begin{itemize}
 \item B: 28.
 \item OB: 2.
 \item WR: 1.
 \item No previous classification: 6.
\end{itemize}

\begin{figure}
\center
\includegraphics[width=0.48\textwidth,viewport=18 48 536 536]{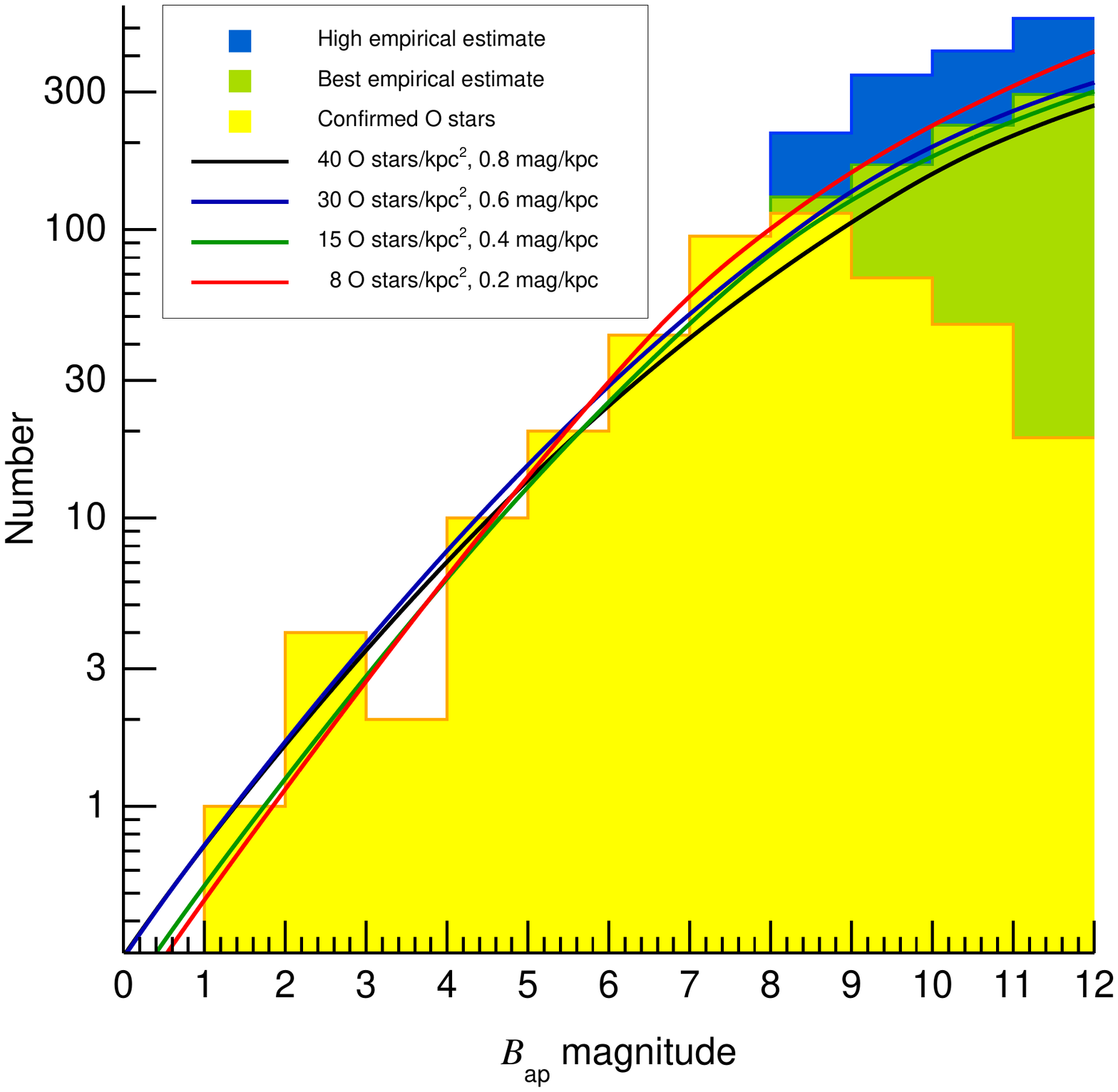} \
\includegraphics[width=0.48\textwidth,viewport=18 48 536 536]{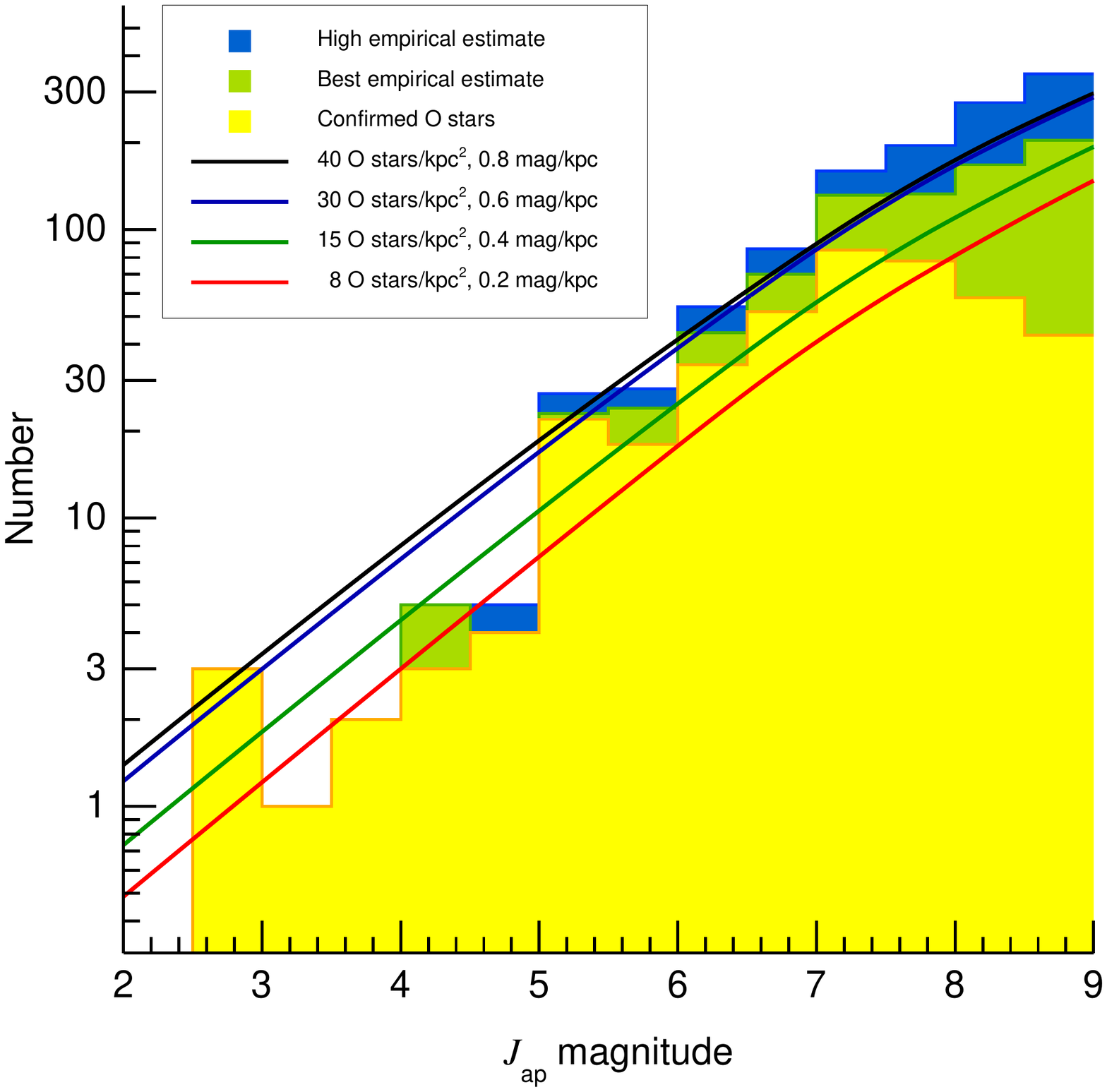}
\caption{Empirical magnitude histograms and toy-model predictions. The histograms ($B_{\rm ap}$, left panel; 
$J_{\rm ap}$, right panel) show the confirmed and estimated (from the unobserved part of the GOSSS sample) 
magnitude distributions for O stars. Predictions for four cases of our toy model with different values of the 
surface density and the extinction per unit distance are also shown.}
\end{figure}

\begin{figure}
\center
\includegraphics[width=0.48\textwidth,viewport=18 48 536 536]{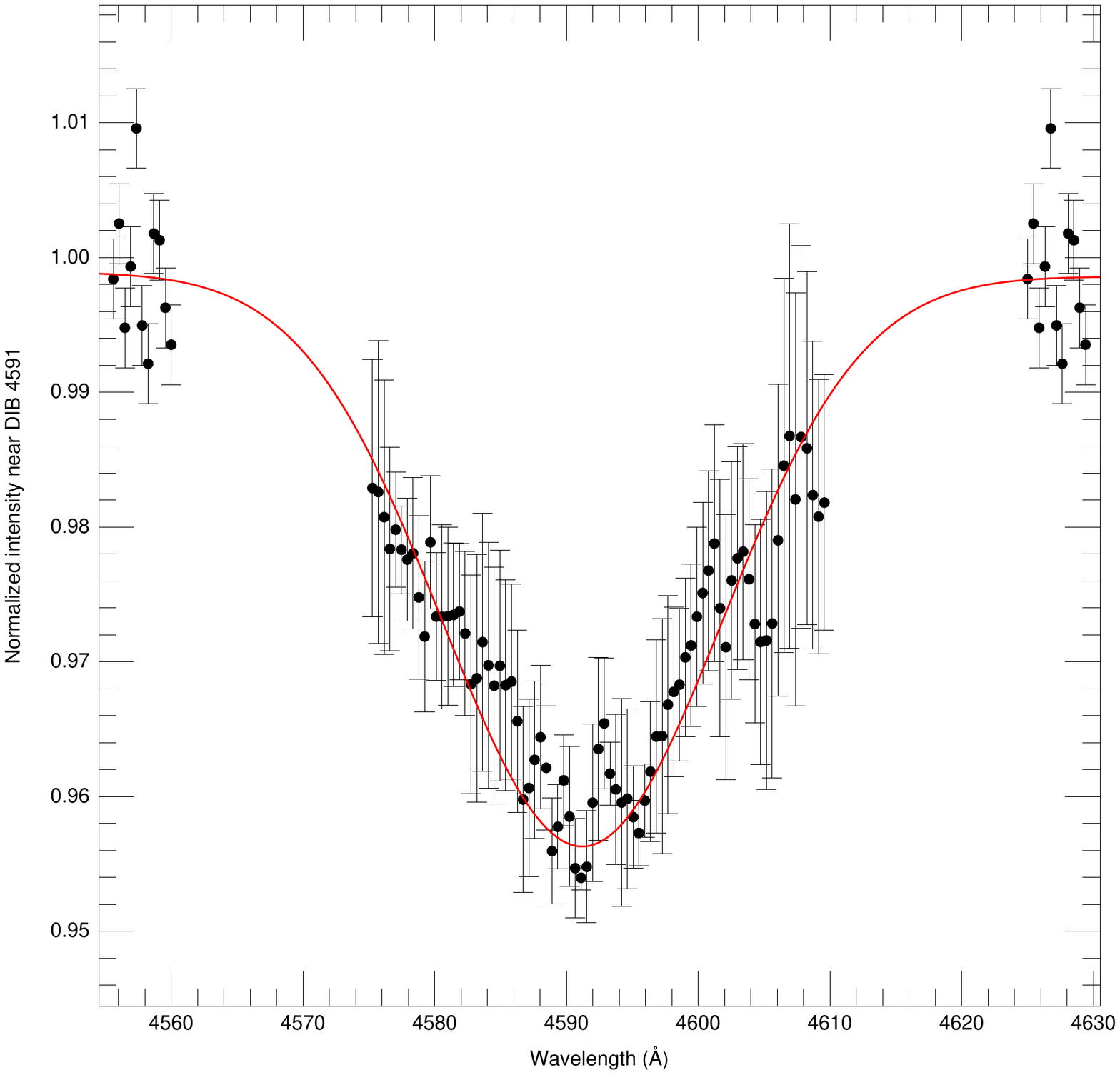} \
\includegraphics[width=0.48\textwidth,viewport=18 48 536 536]{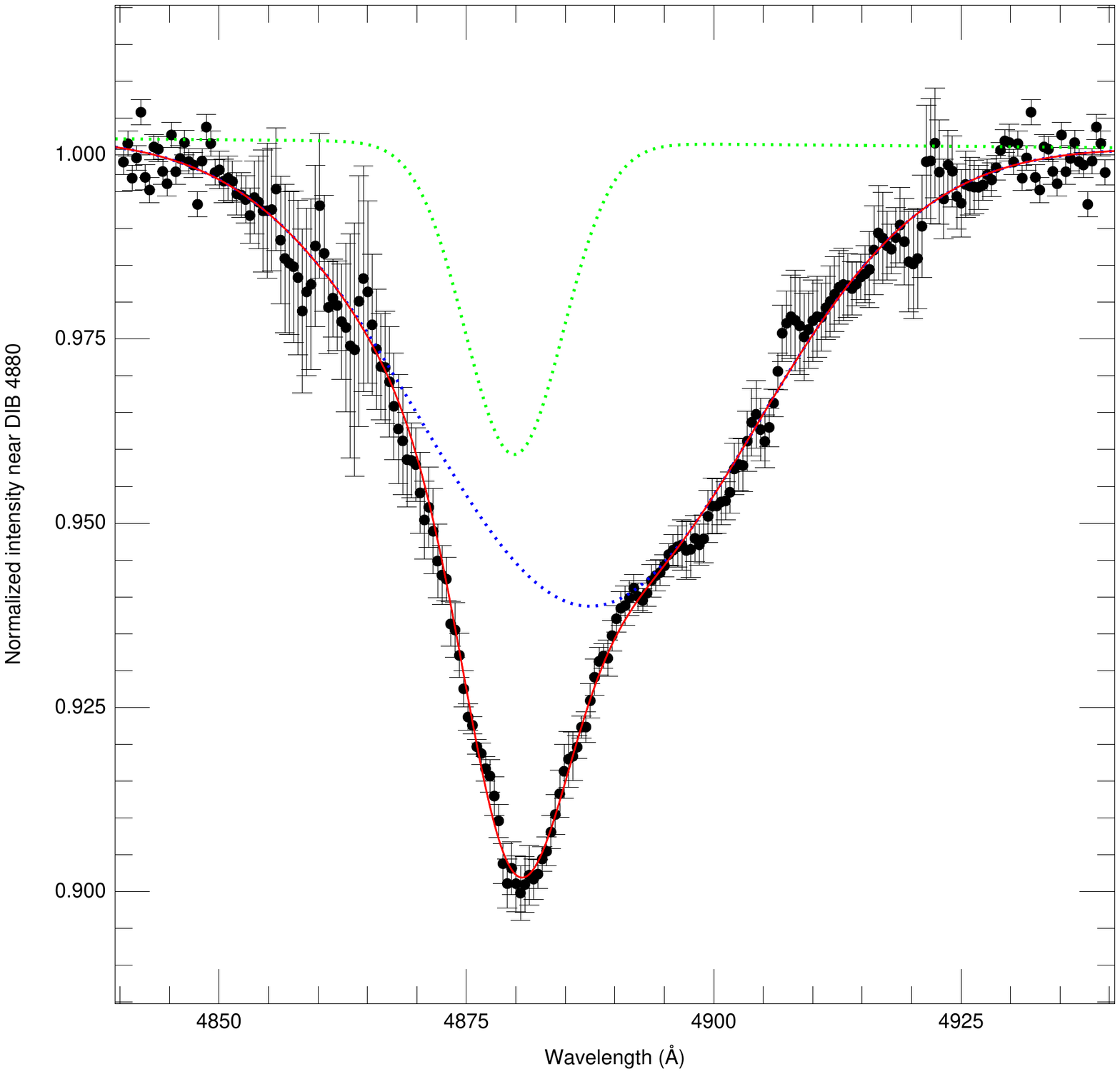}
\caption{Two examples of DIBs observed with GOSSS using the pair method of combining pairs of stars of the 
same spectral type, one with high extinction and one with low extinction. The error bars show the average and 
standard deviation for six (left panel) or seven (right panel) pairs and the red line the Gaussian fit. 
The 4591~\AA~DIB is clearly detected 
for the first time and it has the correct wavelength for the \citet{Ehreetal95} prediction for coronene and 
ovalene cations. In the second case (the region adjacent to H$\beta$) we detect that the broad DIB has two 
components, shown with dotted blue and green lines, respectively. The large error bars around 4860~\AA\ are 
caused by the subtraction of H$\beta$.}
\end{figure}

\subsection{How many O stars are there in the solar neighborhood?}

\begin{itemize}
 \item Toy model to compare observations:
  \begin{itemize}
   \item Constant surface density of O stars in the Galaxy. 
   \item Disk with a 12 kpc radius.  
   \item Sun at 8 kpc from the center.
   \item Constant extinction per unit distance ($A_V$/kpc) and standard extinction law with $R_{5495} = 3.1$.
   \item Kroupa IMF, Geneva non-rotating evolutionary tracks, and constant star formation rate.
  \end{itemize}
 \item Data:
  \begin{itemize}
   \item GOSSS stars already observed and classified.
   \item Resolved O stars counted individually but SB2 O+O systems counted as a single object.
   \item Prediction on the expected number from candidates and previous rates of false positives and false 
         negatives.
  \end{itemize}
 \item Results (Figure~7):
  \begin{itemize}
   \item $B_{\rm ap}$ distribution is consistent with either low surface density + extinction per unit 
         distance or high surface density + extinction per unit distance.
   \item $J_{\rm ap}$ distribution favors the large values of surface density and extinction per unit 
         distance.
   \item The effect of the Local Bubble and our interarm location creates a dearth of NIR-bright 
         ($J_{\rm ap} < 5$) stars. This effect is also seen in the spatial distribution derived from Hipparcos
         parallaxes \citep{Maiz01a,Maizetal08c}.
   \item 30-40 O stars/kpc$^2$ corresponds to 14\,000-18\,000 O stars in the Milky Way if the toy model holds 
         throughout the whole Galaxy.
   \item $A_V$ per unit distance appears to be 0.6-0.8 mag/kpc but patchy extinction may be unseen in the 
         data.
   \item When all effects are considered (radial density gradient in the Galaxy, spectroscopic binaries, patchy
         extinction), the real number of O stars in the Galaxy may be larger by a factor of 2
         or 3 (30\,000-50\,000).
  \end{itemize}
 \item We will recalculate this with a more sophisticated model at the end of the project.
\end{itemize}

\begin{figure}
\center
\includegraphics[width=\textwidth,viewport=0 0 536 170]{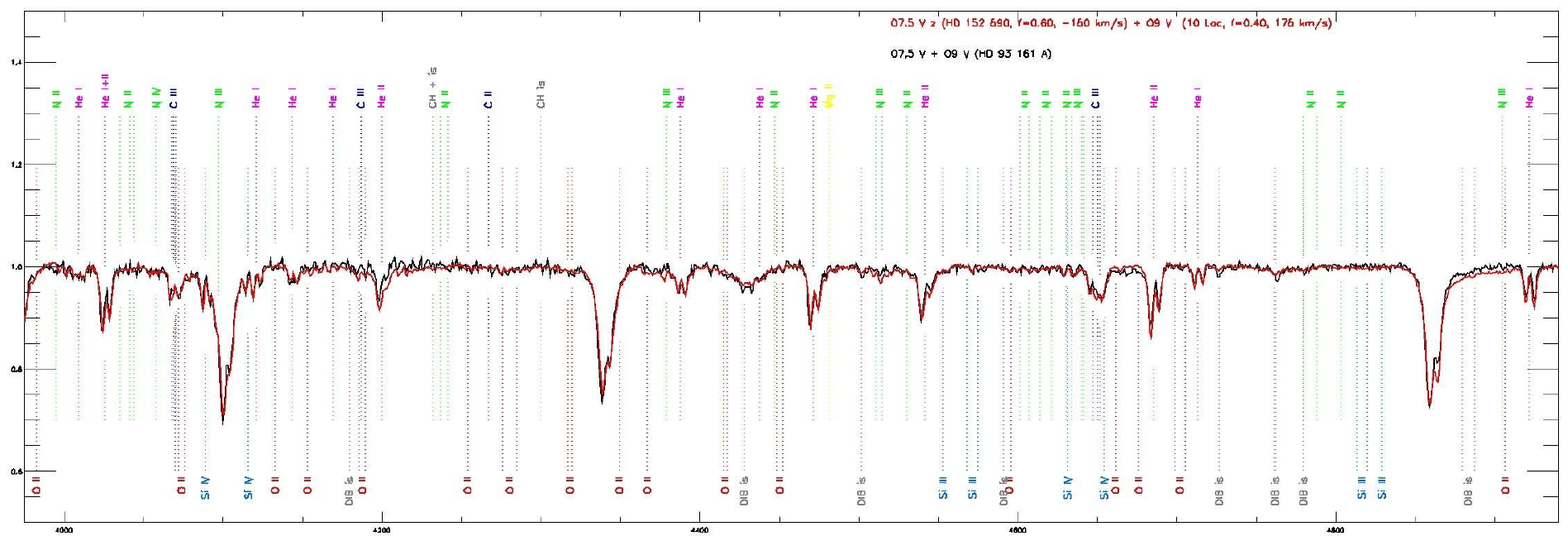} 
\caption{Example of fitting an SB2 system with MGB. Eight parameters can be adjusted: the spectral types, 
luminosity classes, and velocities of both the primary and secondary, the flux fraction of the secondary, and 
the rotation index n. Here HD~93\,161~A (black) is fitted with a combination (red) of 60\% of HD~152\,590 and 
40\% of 10~Lac. See also Figure~3.}
\end{figure}

\section{Diffuse Interstellar Bands with GOSSS}

\begin{itemize}
 \item Why use long-slit intermediate resolution spectroscopy to study DIBs (as opposed to \'echelle 
       spectroscopy)?
 \item Advantages:
  \begin{itemize}
   \item Broad DIBs more clearly seen than in \'echelle data.
   \item Deeper magnitudes reached when all things equal.
   \item Can do two (or even three) stars at a time.
  \end{itemize}
 \item Disadvantages: 
  \begin{itemize}
   \item Worse spectral resolution (e.g. no kinematics or structure in narrow DIBs).
   \item Harder to detect weak narrow DIBs.
   \item Smaller spectral coverage in a single setting.
  \end{itemize}
 \item See Figure~8 and \citet{Maizetal13a} for some results.
\end{itemize}

\section{MGB}

\begin{itemize}
 \item Interactive spectral classification tool for OB stars \citep{Maizetal12}.
 \item Uses GOSSS library of standard stars at $R\sim 2500$.
 \item Includes fitting of SB2 systems (see Figure~9 for an example) and rotation index n.
 \item Modular: other libraries can be used (e.g. $R\sim 4000$).
 \item It will be included in GOSSS-DR2.
\end{itemize}

\small  
%
\section*{Acknowledgments}   
%
Support for this work was provided by [a] the Spanish Government Ministerio de Ciencia e Innovaci\'on through 
grants AYA2007-64052, AYA2007-64712, AYA 2010-17631, AYA 2010-15081, the Ram\'on y Cajal Fellowship program, 
and FEDER funds; [b] the Junta de Andaluc\'{\i}a
grant P08-TIC-4075; [c] NASA through grants GO-10205, GO-10602, and GO-10898 from the Space Telescope 
Science Institute, which is operated by the Association of Universities for Research in Astronomy Inc., under 
NASA contract NAS~5-26555; [d] the Direcci\'on de Investigaci\'on de la Universidad de La Serena (DIULS 
PR09101); [e] the ESO-Government of Chile Joint Committee Postdoctoral Grant; and [f] the Chilean Government 
grants FONDECYT Regular 1120668 and FONDECYT Iniciaci\'on 11121550.
This research has made extensive use of [a] Aladin \citep{Bonnetal00};
[b] the SIMBAD database, operated at CDS, Strasbourg, France; and [c] the Washington Double Star Catalog,
maintained at the U.S. Naval Observatory \citep{Masoetal01}. We would like to thank Miguel Penad\'es Ordaz for
his help in the compilation of data for this work.

\bibliographystyle{aj}
\small
\bibliography{general}

\begin{thebibliography}{}

\bibitem[\protect\citeauthoryear{Barb{\'a} et~al.}{Barb{\'a}
  et~al.}{2010}]{Barbetal10}
Barb{\'a}, R.~H., Gamen, R.~C., Arias, J.~I., Morrell, N.~I.,
  Ma{\'{\i}}z~Apell{\'a}niz, J., Alfaro, E.~J., Walborn, N.~R.,  \& Sota, A.
  2010, in Rev. Mex. Astron. Astrof{\'\i}s. (conference series), Vol.~38, 30

\bibitem[\protect\citeauthoryear{Bonnarel et~al.}{Bonnarel
  et~al.}{2000}]{Bonnetal00}
Bonnarel, F., et~al. 2000, A\&AS, 143, 33

\bibitem[\protect\citeauthoryear{Conti \& Alschuler}{Conti \&
  Alschuler}{1971}]{ContAlsc71}
Conti, P.~S.,  \& Alschuler, W.~R. 1971, ApJ, 170, 325

\bibitem[\protect\citeauthoryear{Ehrenfreund et~al.}{Ehrenfreund
  et~al.}{1995}]{Ehreetal95}
Ehrenfreund, P., Foing, B.~H., D'Hendecourt, L., Jenniskens, P.,  \& Desert,
  F.~X. 1995, A\&A, 299, 213

\bibitem[\protect\citeauthoryear{Feinstein, Moffat, \& Fitzgerald}{Feinstein
  et~al.}{1980}]{Feinetal80}
Feinstein, A., Moffat, A.~F.~J.,  \& Fitzgerald, M.~P. 1980, AJ, 85, 708

\bibitem[\protect\citeauthoryear{Gamen et~al.}{Gamen et~al.}{2012}]{Gameetal12}
Gamen, R., Arias, J.~I., Barb{\'a}, R.~H., Morrell, N.~I., Walborn, N.~R.,
  Sota, A., Ma{\'{\i}}z~Apell{\'a}niz, J.,  \& Alfaro, E.~J. 2012, A\&A, 546,
  A92

\bibitem[\protect\citeauthoryear{Garmany, Conti, \& Chiosi}{Garmany
  et~al.}{1982}]{Garmetal82}
Garmany, C.~D., Conti, P.~S.,  \& Chiosi, C. 1982, ApJ, 263, 777

\bibitem[\protect\citeauthoryear{Garrison, Hiltner, \& Schild}{Garrison
  et~al.}{1977}]{Garretal77}
Garrison, R.~F., Hiltner, W.~A.,  \& Schild, R.~E. 1977, ApJS, 35, 111

\bibitem[\protect\citeauthoryear{Hiltner}{Hiltner}{1956}]{Hilt56}
Hiltner, W.~A. 1956, ApJS, 2, 389

\bibitem[\protect\citeauthoryear{Lesh}{Lesh}{1968}]{Lesh68}
Lesh, J.~R. 1968, ApJS, 17, 371

\bibitem[\protect\citeauthoryear{Ma{\'\i}z~Apell{\'a}niz}{Ma{\'\i}z~Apell{\'a}niz}{2001}]{Maiz01a}
Ma{\'\i}z~Apell{\'a}niz, J. 2001, AJ, 121, 2737

\bibitem[\protect\citeauthoryear{Ma{\'\i}z~Apell{\'a}niz}{Ma{\'\i}z~Apell{\'a}niz}{2004}]{Maiz04c}
Ma{\'\i}z~Apell{\'a}niz, J. 2004, PASP, 116, 859

\bibitem[\protect\citeauthoryear{Ma{\'{\i}}z~Apell{\'a}niz}{Ma{\'{\i}}z~Apell{\'a}niz}{2013}]{Maiz13b}
Ma{\'{\i}}z~Apell{\'a}niz, J. 2013, in Highlights of Spanish Astrophysics VII,
  583

\bibitem[\protect\citeauthoryear{Ma{\'{\i}}z~Apell{\'a}niz, Alfaro, \&
  Sota}{Ma{\'{\i}}z~Apell{\'a}niz et~al.}{2008}]{Maizetal08c}
Ma{\'{\i}}z~Apell{\'a}niz, J., Alfaro, E.~J.,  \& Sota, A. 2008,
  arXiv:0804.2553

\bibitem[\protect\citeauthoryear{Ma{\'{\i}}z~Apell{\'a}niz
  et~al.}{Ma{\'{\i}}z~Apell{\'a}niz et~al.}{2012}]{Maizetal12}
Ma{\'{\i}}z~Apell{\'a}niz, J., et~al. 2012, in Astronomical Society of the
  Pacific Conference Series, Vol. 465, Astronomical Society of the Pacific
  Conference Series, ed. L.~Drissen, C.~Rubert, N.~St-Louis, \& A.~F.~J.
  Moffat, 484

\bibitem[\protect\citeauthoryear{Ma{\'{\i}}z~Apell{\'a}niz
  et~al.}{Ma{\'{\i}}z~Apell{\'a}niz et~al.}{2013}]{Maizetal13a}
Ma{\'{\i}}z~Apell{\'a}niz, J., Sota, A., Barb{\'a}, R.~H., Morrell, N.~I.,
  Pellerin, A., Alfaro, E.~J.,  \& Sim{\'o}n-D{\'{\i}}az, S. 2013,
  arXiv:1305.6163

\bibitem[\protect\citeauthoryear{Ma{\'{\i}}z~Apell{\'a}niz
  et~al.}{Ma{\'{\i}}z~Apell{\'a}niz et~al.}{2011}]{Maizetal11}
Ma{\'{\i}}z~Apell{\'a}niz, J., Sota, A., Walborn, N.~R., Alfaro, E.~J.,
  Barb{\'a}, R.~H., Morrell, N.~I., Gamen, R.~C.,  \& Arias, J.~I. 2011, in
  Highlights of Spanish Astrophysics VI, ed. {M.~R.~Zapatero Osorio, J.~Gorgas,
  J.~Ma{\'{\i}}z Apell{\'a}niz, J.~R.~Pardo, \& A.~Gil de Paz}, 467

\bibitem[\protect\citeauthoryear{Ma{\'\i}z~Apell{\'a}niz
  et~al.}{Ma{\'\i}z~Apell{\'a}niz et~al.}{2004}]{Maizetal04b}
Ma{\'\i}z~Apell{\'a}niz, J., Walborn, N.~R., Galu{\'e}, H.~{\'A}.,  \& Wei,
  L.~H. 2004, ApJS, 151, 103

\bibitem[\protect\citeauthoryear{Mannino \& Humblet}{Mannino \&
  Humblet}{1955}]{MannHumb55}
Mannino, G.,  \& Humblet, J. 1955, Annales d'Astrophysique, 18, 237

\bibitem[\protect\citeauthoryear{Mason et~al.}{Mason et~al.}{2001}]{Masoetal01}
Mason, B.~D., Wycoff, G.~L., Hartkopf, W.~I., Douglass, G.~G.,  \& Worley,
  C.~E. 2001, AJ, 122, 3466

\bibitem[\protect\citeauthoryear{Morgan, Whitford, \& Code}{Morgan
  et~al.}{1953}]{Morgetal53b}
Morgan, W.~W., Whitford, A.~E.,  \& Code, A.~D. 1953, ApJ, 118, 318

\bibitem[\protect\citeauthoryear{Muzzio \& McCarthy}{Muzzio \&
  McCarthy}{1973}]{MuzzMcCa73}
Muzzio, J.~C.,  \& McCarthy, C.~C. 1973, AJ, 78, 924

\bibitem[\protect\citeauthoryear{Schild, Neugebauer, \& Westphal}{Schild
  et~al.}{1971}]{Schietal71}
Schild, R.~E., Neugebauer, G.,  \& Westphal, J.~A. 1971, AJ, 76, 237

\bibitem[\protect\citeauthoryear{Sota et~al.}{Sota et~al.}{2011}]{Sotaetal11a}
Sota, A., Ma{\'{\i}}z~Apell{\'a}niz, J., Walborn, N.~R., Alfaro, E.~J.,
  Barb{\'a}, R.~H., Morrell, N.~I., Gamen, R.~C.,  \& Arias, J.~I. 2011, ApJS,
  193, 24

\bibitem[\protect\citeauthoryear{Sota et~al.}{Sota et~al.}{2008}]{Sotaetal08}
Sota, A., Ma{\'{\i}}z~Apell{\'a}niz, J., Walborn, N.~R.,  \& Shida, R.~Y. 2008,
  in Rev. Mex. Astron. Astrof{\'\i}s. (conference series), Vol.~33, 56

\bibitem[\protect\citeauthoryear{Wade et~al.}{Wade et~al.}{2012}]{Wadeetal12b}
Wade, G.~A., et~al. 2012, MNRAS, 425, 1278

\bibitem[\protect\citeauthoryear{Walborn}{Walborn}{1973}]{Walb73a}
Walborn, N.~R. 1973, AJ, 78, 1067

\bibitem[\protect\citeauthoryear{Walborn et~al.}{Walborn
  et~al.}{2011}]{Walbetal11}
Walborn, N.~R., Ma{\'{\i}}z~Apell{\'a}niz, J., Sota, A., Alfaro, E.~J.,
  Morrell, N.~I., Barb{\'a}, R.~H., Arias, J.~I.,  \& Gamen, R.~C. 2011, AJ,
  142, 150

\bibitem[\protect\citeauthoryear{Walborn et~al.}{Walborn
  et~al.}{2010}]{Walbetal10a}
Walborn, N.~R., Sota, A., Ma{\'{\i}}z~Apell{\'a}niz, J., Alfaro, E.~J.,
  Morrell, N.~I., Barb{\'a}, R.~H., Arias, J.~I.,  \& Gamen, R.~C. 2010, ApJL,
  711, L143

\end{thebibliography}

\end{document}